# Transition between canted antiferromagnetic and spin-polarized ferromagnetic quantum Hall states in graphene on a ferrimagnetic insulator

Yang Li,[1,2]† Mario Amado,[1]† Timo Hyart,[3] Grzegorz P. Mazur,[3] Vetle Risinggård,[4,5] Thomas Wagner,[1,6] Lauren McKenzie-Sell,[1,7] Graham Kimbell,[1] Joerg Wunderlich,[6,8,9] Jacob Linder,[4,5] and Jason W. A. Robinson[1]*

[1]*Department of Materials Science & Metallurgy, University of Cambridge, 27 Charles Babbage Road, Cambridge CB3 0FS, United Kingdom*

[2]*Cambridge Graphene Centre, University of Cambridge, 9 JJ Thomson Avenue, Cambridge CB3 0FA, United Kingdom*

[3] *International Research Centre MagTop, Institute of Physics, Polish Academy of Sciences, Aleja Lotników 32/46, PL-02668 Warsaw, Poland*

[4]*Department of Physics, Norwegian University of Science and Technology, NO-7491 Trondheim, Norway*

[5]*Center for Quantum Spintronics, Department of Physics, Norwegian University of Science and Technology, NO-7491 Trondheim, Norway*

[6]*Hitachi Cambridge Laboratory, Cambridge CB3 0HE, United Kingdom*

[7]*Cavendish Laboratory, University of Cambridge, Cambridge CB3 0HE, United Kingdom*

[8]*Institute of Experimental and Applied Physics, University of Regensburg, Universitätsstrasse 31, 93051 Regensburg, Germany*

[9]*Institute of Physics, Czech Academy of Sciences, Cukrovarnicka 10, 162 00, Praha 6, Czech Republic*

In the quantum Hall regime of graphene, antiferromagnetic and spin-polarized ferromagnetic states at the zeroth Landau level compete, leading to a canted antiferromagnetic state depending on the direction and magnitude of an applied magnetic field. Here, we investigate this transition at 2.7 K in graphene Hall bars that are proximity coupled to the ferrimagnetic insulator $Y_3Fe_5O_{12}$. From nonlocal transport measurements, we demonstrate an induced magnetic exchange field in graphene, which lowers the magnetic field required to modulate the magnetic state in graphene. These results show that a magnetic proximity effect in graphene is an important ingredient for the development of two-dimensional materials in which it is desirable for ordered states of matter to be tunable with relatively small applied magnetic fields (> 6 T).

*Corresponding author E-mail: jjr33@cam.ac.uk

†These authors contributed equally to this work.

Graphene has two inequivalent Dirac cones in the energy band dispersion, which lead to a set of Landau levels with distinct features over conventional two-dimensional electron gases, e.g., in an applied magnetic field ($B$), there exists fourfold degenerate symmetry-broken zero-energy Landau levels with filling factors $v = 0$ and $\pm 1$ [1–3]. These are gate voltage tunable and described by spin and valley degeneracy. Electron-electron and electron-phonon interactions break valley symmetry and determine the magnetic order of the $v = 0$ state. Theory [4–6] and experiment [7–10] indicate that $v = 0$ is an antiferromagnetic (AF) quantum Hall state [6,9] in which the two sublattice spins of graphene align antiparallel. The Zeeman field associated with an in-plane magnetic field ($B_{//}$) favors a spin-polarized ferromagnetic (F) state [9] (that can be also found at $v = \pm 1$ [7]), but in general the AF and F states compete, leading to a canted antiferromagnetic (CAF) $v = 0$ state at half-filled zero-energy Landau level in which the two sublattice spins tilt out-of-plane. The spin-direction in the CAF state depends on the sum of the spin components parallel (preferred by the Zeeman field, responsible for the F state) and perpendicular to $B$ (preferred by the electron-electron Coulomb interactions responsible for the valley anisotropy and leading to the AF state). In the AF state, there are gapped edge modes while the F state supports gapless counter-propagating edge modes [5,7,11]. Therefore, in the CAF state the energy gap of the edge modes is tunable with the direction and magnitude of $B$ [5].

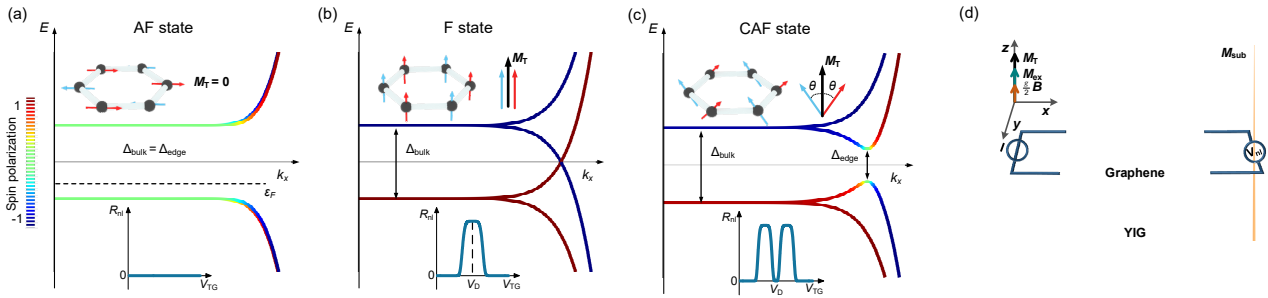

**FIG. 1.** Energy spectra for the (a) AF, (b) F and (c) CAF states in graphene, which arise depending on the total magnetic field ($M_T$) applied to graphene and the angle ($\theta$) between $M_T$ and the sublattice spins. Color scale bar shows -1 (spin-direction antiparallel to $M_T$) to 1 (spin-direction parallel to $M_T$). Top insets in (a)-(c): Schematic diagrams illustrating the sublattice spins in graphene (left) which make an angle $\theta$ with respect to $M_T$ (right). Bottom insets in (a)-(c): $R_{nl}$ vs $V_{TG}$ for AF, F, and CAF states near $V_D$. The AF state (a) forms when $M_T$ is zero (sublattice spins are antiparallel). The F state (b) forms when $M_T$ is larger than a critical value (determined by the Coulomb interaction) and the sublattice spins are parallel to $M_T$. The CAF state (c) is a mixture of AF and F states and forms when $M_T$ is nonzero, but smaller than a critical value (sublattice spins are noncollinear to $M_T$). (d) Schematic diagram of a graphene Hall bar on YIG in which magnetization ($M_{sub}$) induces a nonzero $M_{ex}$ which adds to the Zeeman field ($\frac{g}{2}B$).

Edge modes associated with CAF and F states have been detected in graphene Hall bars through nonlocal measurements with a transition between F and CAF states occurring around 15 T – 30 T [9]. In the ballistic limit, the nonlocal resistance ($R_{nl}$) is quantized and dependent on the Hall bar geometry [12]. In the diffusive limit, $R_{nl}$ is not quantized but shows different behaviors with gate voltage ($V_{TG}$) in the AF, CAF and F states [Figs. 1(a)-1(c)]: the AF state does not support edge modes meaning $R_{nl}$ = 0, but the CAF (F) is gapped (gapless) and $R_{nl}$ shows a double peak (single peak) around the Dirac point ($V_D$).

Transitions between CAF and F (or AF) states could be achieved in lower applied magnetic fields if graphene has an intrinsic magnetic exchange field ($M_{ex}$). By placing graphene on an insulating magnetic substrate, a hybridization of the $\pi$ orbitals in graphene with the substrate can theoretically induce a magnetic exchange field of hundreds of Tesla [13–18]. The magnitude of the total magnetic field ($M_T = \|M_T\|$) applied to graphene is then related to $M_T = \frac{g}{2}B + M_{ex}$, where $g$ is gyromagnetic ratio and $\frac{g}{2}B$ is the Zeeman field. $M_{ex}$ is parallel to $B$. A 14-T magnetic exchange field was recently estimated in graphene on EuS [14] and an anomalous Hall effect in graphene on $Y_3Fe_5O_{12}$ (YIG) showed evidence for an induced magnetic exchange field in graphene [19]. In [14,19], transitions between CAF and F (or AF) states were not investigated.

Here we report transitions between CAF and F states in hexagonal boron nitride (hBN) covered graphene Hall bars on YIG. These are investigated through gate-voltage-dependent nonlocal transport measurements below 9 K. The magnetic state and energy gap of the edge modes in graphene are tunable by varying the magnitude (> 6 T) and direction of an applied magnetic field (**B**). The tunable energy gap is important from a fundamental viewpoint, as it separates quantum states with distinct magnetic ordering in graphene, and implies potential relevance for applications requiring tunable band gap, such as photodetectors.

YIG has a Curie temperature of 550 K, a band gap of 2.84 eV and an electrical resistivity of $10^{12}$ $\Omega$ cm. Moreover, it is chemically stable in the air, which minimizes surface degradation during Hall bar fabrication. Atomically flat (1 1 0) YIG (84-nm-thick) is prepared by pulsed laser deposition onto gadolinium gallium garnet (GGG) [Fig. 2(a) and bottom inset] with a bulk magnetization of 144 emu cm$^{-3}$ (see Fig. S3 in [21]). Hall bars are fabricated in several steps, in which graphene is exfoliated from graphite and transferred onto YIG. The graphene is covered with a 20-to-50-nm-thick layer of hBN and electron beam lithography defines Cr/Au side-contacts [21]. In this paper, we report two hBN/graphene Hall bars on YIG, which show field-effect mobility ($\mu$) of around 12,000 cm$^2$V$^{-1}$s$^{-1}$ (Device I) and 10,000 cm$^2$V$^{-1}$s$^{-1}$ (Device II) at 9 K. Control Hall bar of hBN/graphene/AlO$_x$/YIG ($\mu \approx$ 15,000 cm$^2$V$^{-1}$s$^{-1}$ at 9 K) is investigated in which graphene is decoupled from YIG with a 6-nm-thick AlO$_x$ layer. Raman spectroscopy is performed on the graphene prior to and following transfer onto YIG or AlO$_x$ [top inset in Fig. 2(a)] and shows no evidence for defects in graphene.

Figure 2(b) (left inset) shows a representative hBN/graphene Hall bar on YIG prior to top-gate electrode deposition. Resistance is measured using lock-in amplifiers [21]. For local transport, $I_{9,10}$ indicates current flowing between contacts 9 and 10 and a local voltage $V_{3,5}$ is measured between contacts 3 and 5, giving $R_{xx} = V_{3,5}/I_{9,10}$. The nonlocal voltage is probed away from the current path (e.g. $R_{nl} = R_{34,56} = V_{5,6}/I_{3,4}$).

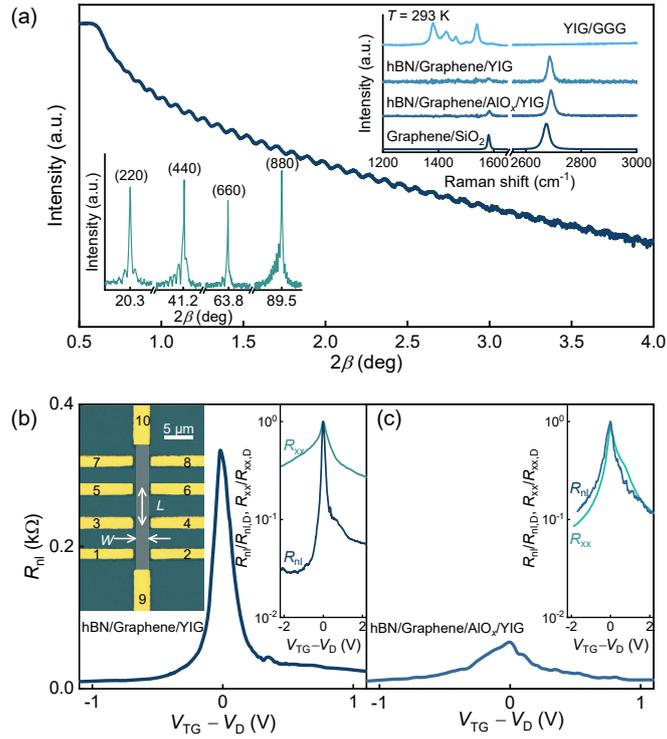

**FIG. 2.** (a) X-ray reflectivity of YIG (84-nm-thick with a roughness of 0.14 nm) on GGG. Upper inset: Raman spectra at 293 K for different structures (labelled) in which the background Raman spectra from hBN and YIG/GGG are subtracted. The G-peak (≈1580 cm$^{-1}$) and 2D-peak (≈2700 cm$^{-1}$) positions of graphene on different substrates vary due to different doping levels [20]. Lower inset: X-ray diffraction trace showing single phase (1 1 0) YIG. (b)-(c) $R_{nl}$ vs $V_{TG} - V_D$ at 9 K for an hBN/graphene Hall bar on YIG (Device I) and a control Hall bar (labelled) with insets (right) showing $R_{xx}/R_{xx,D}$ and $R_{nl}/R_{nl,D}$ vs $V_{TG} - V_D$ for the same Hall bar in zero magnetic field. Left inset of (b): False color optical image of an hBN/graphene Hall bar on YIG.

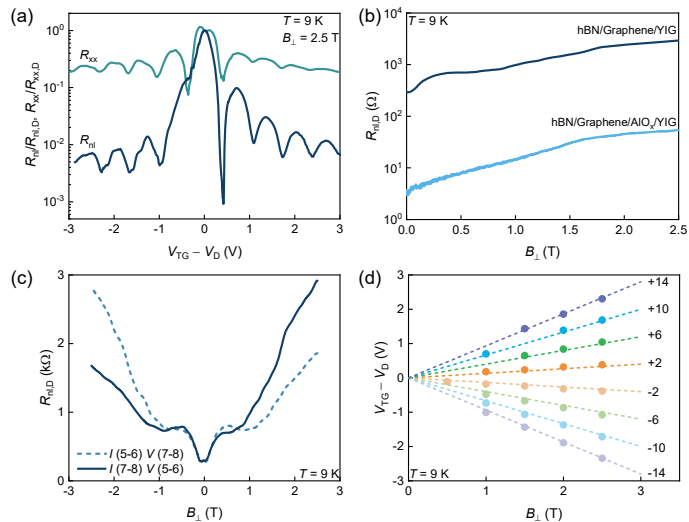

**FIG. 3.** (a) Gate-voltage-dependence of $R_{xx}/R_{xx,D}$ and $R_{nl}/R_{nl,D}$ with $B_\perp$ = 2.5 T. (b) $R_{nl,D}$ with $B_\perp$ for Device I compared to the hBN/graphene/AlO$_x$/YIG control Hall bar. (c) $R_{nl,D}$ vs $B_\perp$ for reverse electrical connections showing that the Onsager relation is obeyed in Device I. (d) Landau level fan diagram where the dashed lines are calculated fitting results. Filling factors are shown beside the dashed lines. All data are recorded at 9 K.

We first discuss transport properties in zero magnetic field for Device I at 9 K. Figure 2(b) shows a peak in $R_{nl}$ at $V_D$. By normalizing $R_{xx}$ and $R_{nl}$ to their maximum values at $V_D$ ($R_{xx,D}$ or $R_{nl,D}$), we see that $R_{nl}/R_{nl,D}$ is an order of magnitude smaller than $R_{xx}/R_{xx,D}$ and the peak in $R_{nl}$ is sharper than $R_{xx}$ [right inset in Fig. 2(b)]. The peak in $R_{nl}$ (≈380 Ω) at $V_D$ may indicate a contribution from the spin Hall [22] or Zeeman spin Hall effects [14]. However, $R_{xx}$ shows a negative magnetoresistance (weak localization) (see Fig. S5(b) in [21]) at 2.7 K suggesting that Rashba spin-orbit coupling is not strong at the graphene/YIG interface, and meaning that the spin Hall effect is unlikely to dominate $R_{nl}$. The YIG has a small remanent out-of-plane magnetic moment (see Fig. S3(d) in [21]) which may support the Zeeman spin Hall effect. We note that ohmic, Joule heating and Ettingshausen contributions to $R_{nl}$ are negligible [21]. Equivalent measurements on an hBN/graphene/AlO$_x$/YIG control Hall bar [Fig. 2(c)] show a reduced $R_{nl}$ at $V_D$ of ≈65 Ω at 9 K, which is dominated by the ohmic effect. This suggests that in zero magnetic field, $R_{nl,D}$ of Device I is due to a coupling between graphene and YIG.

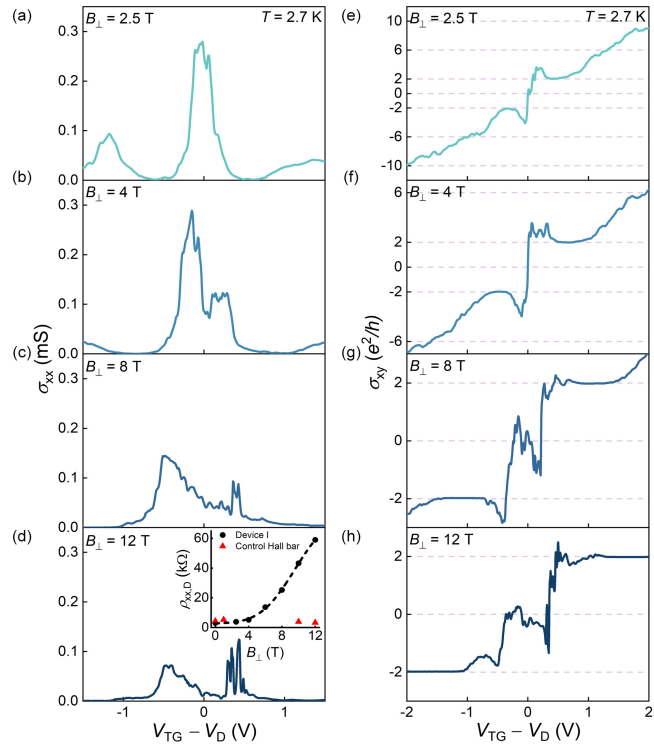

**FIG. 4.** (a)-(d) Gate-voltage-dependence of the longitudinal conductance ($\sigma_{xx}$) for increasing $B_\perp$ (labelled) and (e)-(h) the corresponding Hall conductance ($\sigma_{xy}$) over the same magnetic field range. Inset of (d) shows the longitudinal resistivity ($\rho_{xx,D}$) at $V_D$ vs $B_\perp$ for Device I and the hBN/graphene/AlO$_x$/YIG control Hall bar. All data are recorded at 2.7 K.

In Fig. 3(a) we show gate-voltage-dependent Shubnikov-de Haas oscillations in $R_{xx}$ and $R_{nl}$ (normalized to values at $V_D$) for Device I with an out-of-plane magnetic field of 2.5 T. The ratios $R_{nl}/R_{nl,D}$ and $R_{xx}/R_{xx,D}$ show different trends with gate voltage with $R_{nl}$ decreasing faster than $R_{xx}$. Furthermore, the value of $R_{nl}$ is a factor of 50 larger than in the hBN/graphene/AlO$_x$/YIG control Hall bar [Fig. 3(b)]. These observations in conjunction with the fact that the Onsager relation $R_{56,78}(B_\perp) = R_{78,56}(-B_\perp) \neq R_{78,56}(B_\perp)$ for Device I holds [Fig. 3(c)], demonstrate a contribution to $R_{nl}$ from the Zeeman spin Hall effect [14,23] due to an induced magnetic exchange field. Shubnikov-de Haas oscillations are observed for $B_\perp \geq 1$ T (see Fig. S6 in [21]) showing Landau levels at filling factors $v = 4(N+1/2)$ where $N = 0, \pm 1, \pm 2...$ [Fig. 3(d)]. In addition, the induced magnetic exchange field manifests through the appearance of anomalous Hall effect in Fig. S9 [21].

For large $B_\perp$, quantum Hall plateaus at $v = 0$ and $\pm 1$ may become visible. The $v = 0$ state at the half-filled zeroth Landau level should show a minimum in longitudinal conductance ($\sigma_{xx}$) while the other filling factors at a quarter and three-quarter occupancy are at a maximum. In Figs. 4(a)-4(d) we show the dependence of $\sigma_{xx}$ on gate voltage for increasing values of $B_\perp$: in 4 T a minimum in $\sigma_{xx}$ is visible at $V_D$ and approaches zero in 12 T. Over the same magnetic field range at $V_D$, $\rho_{xx,D}$ rapidly increases [inset of Fig. 4(d)] indicating a transition to a gapped bulk state. Simultaneously the Hall conductance ($\sigma_{xy}$) tends to be a plateau establishing the $v = 0$ state. We note that, equivalent measurements on the hBN/graphene/AlO$_x$/YIG control Hall bar do not show these trends [21] indicating that a coupling between graphene and YIG reduces $B_\perp$ to achieve the $v = 0$ state.

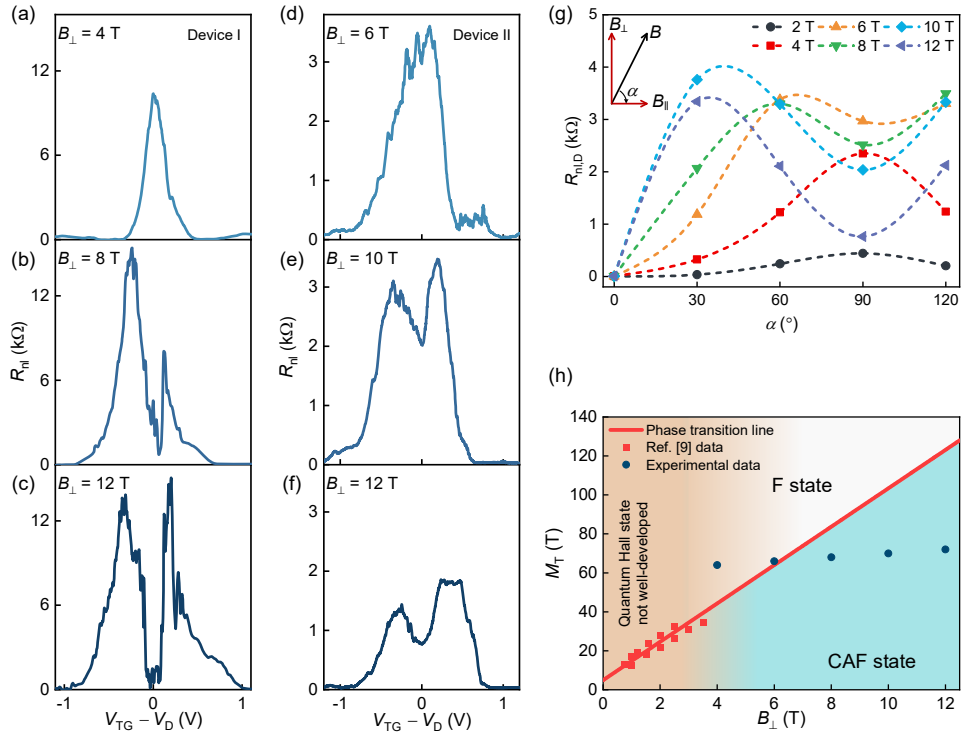

**FIG. 5.** (a)-(f) Gate-voltage-dependence of $R_{nl}$ for different values of $B_\perp$ (labelled) for Devices I (a)-(c) and II (d)-(f). (g) $R_{nl,D}$ vs $\alpha$ with $B$ from 2 T to 12 T for Device II. Dashed lines are a guide to the eye. A ±5° operational error due to manual rotation of the sample holder leads to the small asymmetry in $R_{nl,D}$ at $\alpha = 60°$ and $120°$. (h) Magnetic phase diagram ($M_T$ vs $B_\perp$) for graphene in which the solid (red) line $M_T \approx 9.9 B_\perp + 4.9$ is calculated from [9] using the extracted data in red as explained in the main text and Ref. [21]. The blue data represent the estimated phases for Devices I and II with $B_\perp$ only. For small $B_\perp$, the quantum Hall state is not well-developed. By increasing $B_\perp$, there exists a transition between F and CAF state. For reasonably small $B_\perp$ and large $M_T$, the F state is realized, whereas by increasing the ratio of $B_\perp/M_T$, the CAF state becomes energetically favored. All data are recorded at 2.7 K except the data from [9], which are at 300 mK.

The $v$ = 0 state in Fig. 4 could be a F or a CAF state. These are distinguishable from the gate-voltage-dependence of $R_{nl}$ with $B_\perp$ as shown in Figs. 5(a)-5(c) for Device I and Figs. 5(d)-5(f) for Device II; the transition from a single to a double peak in $R_{nl}$ with gate voltage suggests that the $v$ = 0 state is associated with a transition from F to CAF state. Although consistent with theory [5], the transition occurs in graphene at lower values of $B_\perp$ than without YIG (> 15 T in [9]). Furthermore, the decrease in $R_{nl}$ at $V_D$ with increasing $B_\perp$ suggests an increase in the edge gap and the angle between $M_T$ and the sublattice spins [21]. This angle increases with $B_\perp$ because the valley anisotropy energy (resulted from electron-electron interactions, which leads to an AF state) increases faster than the Zeeman energy as discussed in [21]. To test this hypothesis, we rotated Device II from $\alpha$ = 90° to $\alpha$ = 0° using magnetic fields of 2 to 12 T, where $\alpha$ is the angle between the Hall bar surface and $B$ [Fig. 5(g)]. This rotation partially (6 T < $\|B\|$ < 12 T) or fully ($\|B\|$ = 12 T) transforms the CAF to F state. As the fixed $\|B\|$ rotates in-plane, $B_\perp$ decreases and the sublattice spins align to $M_T$ reducing the edge gap and increasing $R_{nl,D}$.

Transitions between CAF and F states were investigated in [9] using hBN/graphene Hall bars with rotating $B$ up to 35 T at 300 mK. In that work the graphene was not in contact with a magnetic substrate meaning $\|M_{ex}\| = 0$ and thus there is only a Zeeman field. By extracting the average values of $M_T$ vs $B_\perp$ in [9], we calculate a phase transition line of $M_T \approx 9.9 B_\perp + 4.9$ which separates the CAF and F states as shown in Fig. 5(h) and explained in [21]. For Devices I and II [Figs. 5(a)-5(f)], transitions between the CAF and F states occur from $B_\perp$ > 6 T. By comparing our results with Ref. [9] and using $M_T = \frac{g}{2} B + M_{ex}$, we estimate $M_{ex}$ in graphene to be of the order 60 T due to the magnetic proximity effect. This estimate assumes that $\|M_{ex}\|$ is independent of $B$ as long as $B$ is enough to fully magnetize YIG (which is the case in our experiment) [21].

In conclusion, we have demonstrated that by proximity-inducing a magnetic exchange field in graphene on a ferrimagnetic substrate, transitions between CAF and F states can be achieved with relatively low applied magnetic fields (> 6 T) at 2.7 K. This achievement is important for the development of two-dimensional materials with magnetic-field-tunable ordered states of matter.

**Acknowledgements**

The work was funded by the Royal Society and the Engineering and Physical Sciences Research Council (EPSRC, Grant No. EP/P026311/1). Y.L. was supported by the China Scholarship Council (CSC) Cambridge International Scholarship. J.L. and V.R. received funding from the Outstanding Academic Fellows Program at the NTNU, the NV-Faculty, and the Research Council of Norway (Grants No. 216700 and No. 240806) and funding for the Centre of Excellence QuSpin (Grant No. 262633). T.H. and G.P.M. were supported by the Foundation for Polish Science through the IRA Programme cofinanced by EU within SG OP. G.P.M. was supported by the National Science Center (Poland) through ETIUDA fellowship (Grant No. UMO-2017/24/T/ST3/00501).

# Supplementary Materials for

**Transition between canted antiferromagnetic and spin-polarized ferromagnetic quantum Hall states in graphene on a ferrimagnetic insulator**

Yang Li,[1,2,*] Mario Amado,[1,*] Timo Hyart,[3] Grzegorz P. Mazur,[3] Vetle Risinggård,[4,5] Thomas Wagner,[1,6] Lauren McKenzie-Sell,[1,7] Graham Kimbell,[1] Joerg Wunderlich,[6,8,9] Jacob Linder,[4,5] and Jason W. A. Robinson[1,†]

[1] *Department of Materials Science & Metallurgy, University of Cambridge, 27 Charles Babbage Road, Cambridge CB3 0FS, United Kingdom*

[2] *Cambridge Graphene Centre, University of Cambridge, 9 JJ Thomson Avenue, Cambridge CB3 0FA, United Kingdom*

[3] *International Research Centre MagTop, Institute of Physics, Polish Academy of Sciences, Aleja Lotników 32/46, PL-02668 Warsaw, Poland*

[4] *Department of Physics, Norwegian University of Science and Technology, NO-7491 Trondheim, Norway*

[5] *Center for Quantum Spintronics, Department of Physics, Norwegian University of Science and Technology, NO-7491 Trondheim, Norway*

[6] *Hitachi Cambridge Laboratory, Cambridge CB3 0HE, United Kingdom*

[7] *Cavendish Laboratory, University of Cambridge, Cambridge CB3 0HE, United Kingdom*

[8] *Institute of Experimental and Applied Physics, University of Regensburg, Universitätsstrasse 31, 93051 Regensburg, Germany*

[9] *Institute of Physics, Czech Academy of Sciences, Cukrovarnicka 10, 162 00, Praha 6, Czech Republic*

*These authors contributed to this work.

†Corresponding author: jjr33@cam.ac.uk

## S1. Growth of YIG

(1 1 0) epitaxial YIG is grown from a stoichiometric target by pulse laser deposition (KrF laser, wavelength $\lambda$ = 248 nm) at 750 °C in flowing $O_2$ at 0.12 mbar with a pulse fluence of 2.2 J cm$^{-2}$ for 40 minutes and 4 Hz repetition rate onto lattice-matched 5 × 5 mm$^2$ (1 1 0) GGG. The films are annealed at 850 °C for 2 hours in 0.5 mbar of static $O_2$, and subsequently cooled at a rate of 5°C min$^{-1}$.

## S2. Hall bar fabrication

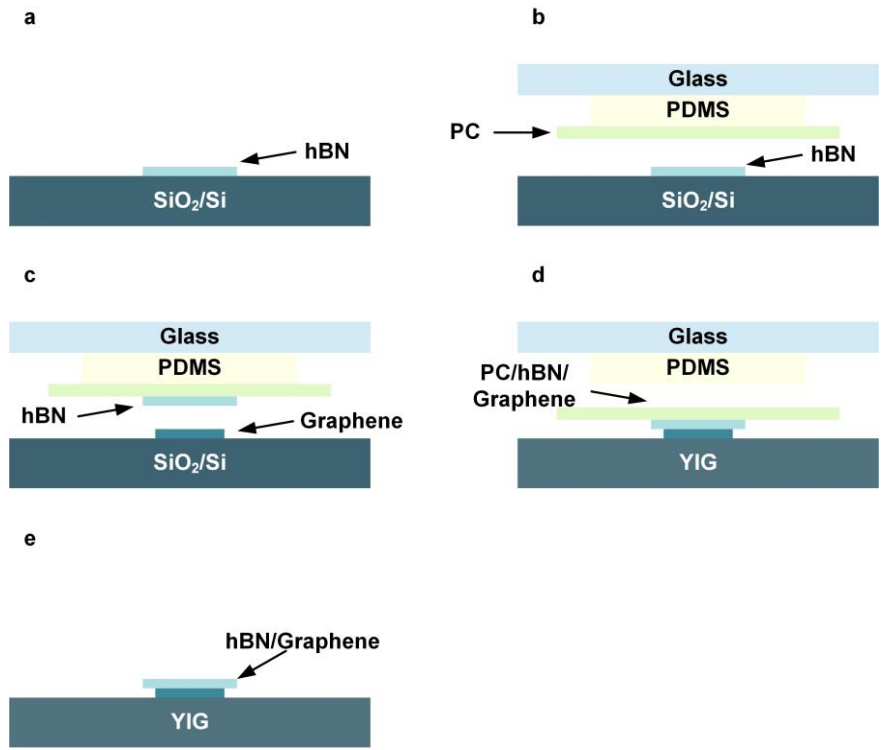

FIG. S1. Transfer procedures for hBN/graphene on YIG. (a) Exfoliation of hBN flakes on SiO$_2$/Si. (b) Transfer hBN onto the stamp. (c) Transfer graphene onto hBN/stamp. (d) Transfer polycarbonate (PC)/hBN/graphene onto YIG. (e) Remove PC film.

Graphene is prepared by mechanical cleavage from high purity graphite and is transferred onto SiO$_2$/Si using pre-fabricated alignment markers. Thin hBN flakes (20-50 nm, confirmed by atomic force microscopy, purchased from HQ Graphene) are prepared by mechanical cleavage from hBN crystals. A stamp is prepared for the transfer process of graphene and hBN, which includes three layers: the first is a piece of thin transparent glass; the second, a transparent and flexible polydimethylsiloxane (PDMS) film, which has two adhesive sides; the third is a thin polycarbonate (PC) film. The selected hBN flake is picked up by a transfer system, which includes stamp, micromanipulator, hot plate and optical microscope. The hBN flake on the stamp peels off the selected graphene on SiO$_2$/Si and transfers graphene from SiO$_2$/Si to YIG. After the stamp touches the YIG substrate, the hot plate is set to 180°C to melt the PC film and then the mask is lifted.

hBN/graphene is released from the PDMS on the glass. Finally, the PC film on the hBN/graphene is dissolved by chloroform. The whole transfer process is illustrated in Fig. S1.

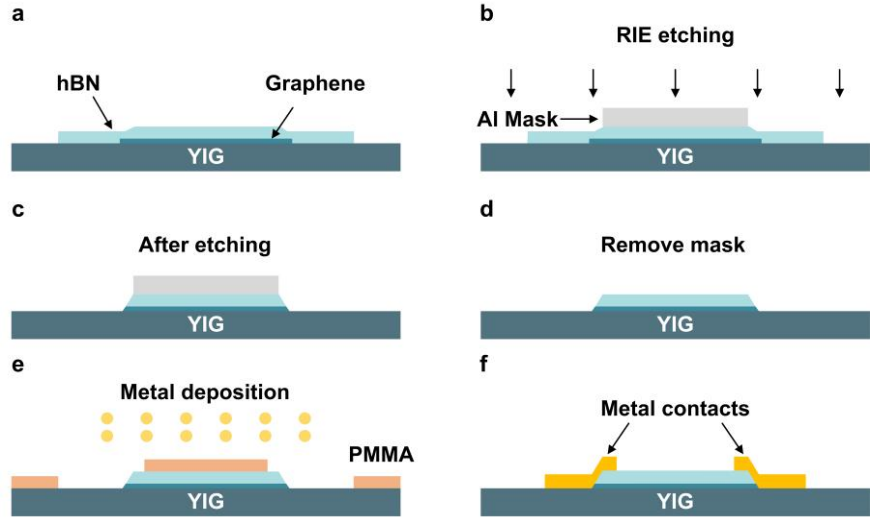

FIG. S2. Side-contacts fabrication procedures for hBN/graphene Hall bar on YIG. (a) hBN/graphene/YIG structure. (b) Reactive ion etching of hBN/graphene on YIG with metal mask. (c) Al/hBN/graphene/YIG structure after etching. (d) hBN/graphene Hall bar pattern. (e) Side-contacts deposition on hBN/graphene/YIG with PMMA mask. (f) The device structure of hBN/graphene Hall bar on YIG.

Hall bars are fabricated by electron beam lithography (EBL) as shown in Fig. S2. A 30-nm-thick Al mask layer is patterned by EBL and deposited by electron beam evaporator. The hBN/graphene is shaped into Hall bar using reactive ion etcher with Al mask. Then the sample is rinsed with AZ 326 MIF developer to remove the Al mask. A double-layer PMMA resist is used to pattern the contacts on the hBN/graphene with EBL. 10-nm-thick Cr and 70-nm-thick Au films are deposited by electron beam evaporation to define side-contacts. The dielectric layer for the top-gate is amorphous $AlO_x$ (40-nm-thick) prepared by atomic layer deposition with trimethylaluminum (TMA) and $H_2O$ as precursors at 120 °C. The top-gate Cr/Au contact is prepared by EBL and electron beam evaporator.

## S3. Characterization of YIG and graphene/YIG

The magnetic properties of YIG are assessed through magnetization ($M$) vs magnetic field ($B$) measurements (Fig. S3). The in-plane $M(B)$ loop is strongly anisotropic with an easy axis coercivity of ≈0.3 mT and a volume magnetization at saturation of 144 emu $cm^{-3}$. The maximum in-plane (out-of-plane) magnetic field required to fully magnetize the YIG is 0.7 mT (200 mT).

To investigate structural and electronic homogeneity of the graphene on YIG, Raman spectra maps are measured at 293 K [Figs. S4(a)-S4(b)]. The positions of the 2D-peak [Pos(2D)] are in the range of 2680-2700 $cm^{-1}$, and the full width at half maximum (FWHM) of the 2D-peak are typically in the range of 20-30 $cm^{-1}$.

From the Raman spectra maps, the homogeneous area [2690 cm$^{-1}$ ≤ Pos(2D) ≤ 2700 cm$^{-1}$ and 20 cm$^{-1}$ ≤ FWHM ≤ 25 cm$^{-1}$] is selected and used to fabricate device [Fig. S4(c)].

The quality of the Hall bars is characterized through Hall-effect and field-effect mobility. The Hall-effect mobility of hBN/graphene Hall bars on YIG can be tuned by the top-gate voltage (from −5 V to 1 V with a leakage current of around 2×10$^{-11}$ A) up to 50,000 cm$^2$ V$^{-1}$ s$^{-1}$ with 5×10$^{10}$ cm$^{-2}$ carrier density at 9 K [Fig. S4(d)] and the field-effect mobility in the 3,000 – 12,000 cm$^2$ V$^{-1}$ s$^{-1}$ range at 9 K [Figs. S4(e)-S4(f)] and 4,000 – 40,000 cm$^2$ V$^{-1}$ s$^{-1}$ at 2.7 K [Fig. S4(g)] with carrier density between 10$^{11}$ and 10$^{12}$ cm$^{-2}$.

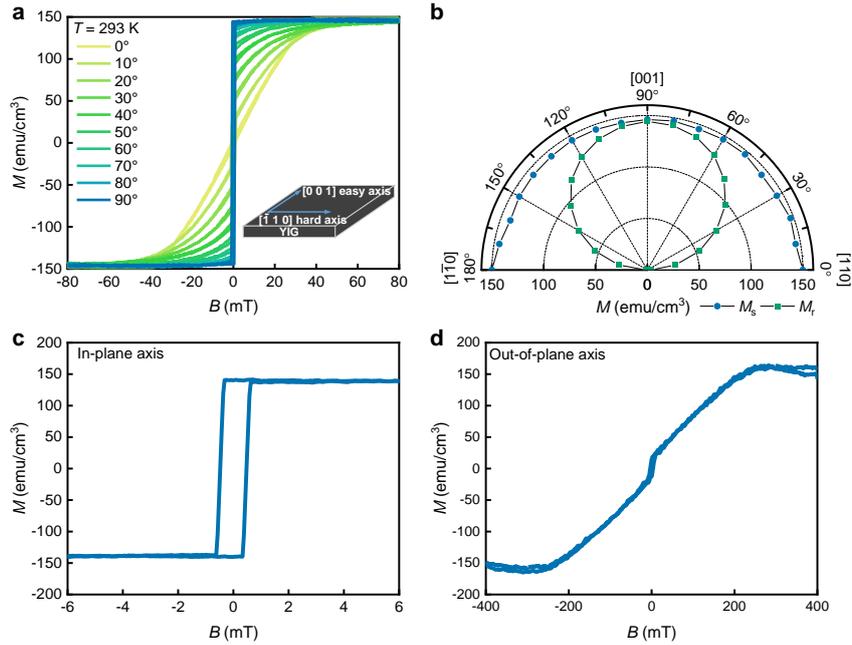

FIG. S3. Magnetic properties of (1 1 0) YIG at 293 K. (a) Magnetization ($M$) vs $B$ hysteresis loops for different in-plane magnetic field directions. At 0°, $B$ is parallel to the hard axis $[\bar{1}10]$, while for 90° along the easy axis [0 0 1]. Inset shows in-plane easy and hard axis directions on (1 1 0) YIG. (b) Stoner plot shows constant saturation magnetization ($M_s$) and the variation of the remanent magnetization ($M_r$) on γ where γ = 0° is parallel to $[\bar{1}10]$. (c)-(d) Low fields $M(B)$ along the in-plane easy axis (c) and out-of-plane hard axis (d).

## S4. Transport measurement setup

Transport measurements are performed using lock-in amplifiers at low frequency (7 Hz) with an excitation current of 50 nA at 2.7 K and 100 nA at 9 K as a function of magnetic field (0-12 T), top-gate voltage and temperature ($T$ > 2.5 K). A series resistance of 10 MΩ or 100 MΩ is introduced to maintain a constant current condition that is confirmed by the signal from the lock-in amplifier which measures the current through a 10 kΩ series resistor. For local measurements in Fig. S4(h), a current source (e.g. $I_{9,10} = V_{10k\Omega}/R_{10k\Omega}$) is applied between the contacts (e.g. 9 and 10), the measured voltage between the contacts (1 and 2) is Hall voltage ($V_{1,2}$) and between the contacts (2 and 4) is longitudinal voltage ($V_{2,4}$). The Hall resistance ($R_{xy}$) is calculated by $R_{xy} = V_{1,2}/I_{9,10}$, and longitudinal resistance $R_{xx} = V_{2,4}/I_{9,10}$. For the nonlocal measurement in Fig. S4(i), a current source ($I_{3,4} = V_{10k\Omega}/R_{10k\Omega}$) is applied between the contacts (3 and 4), the measured voltage between the contacts (1 and 2) is nonlocal voltage ($V_{1,2}$) and is often converted to nonlocal resistance ($R_{nl}$) by dividing the injection current ($R_{nl} = R_{34,12} = V_{1,2}/I_{3,4}$).

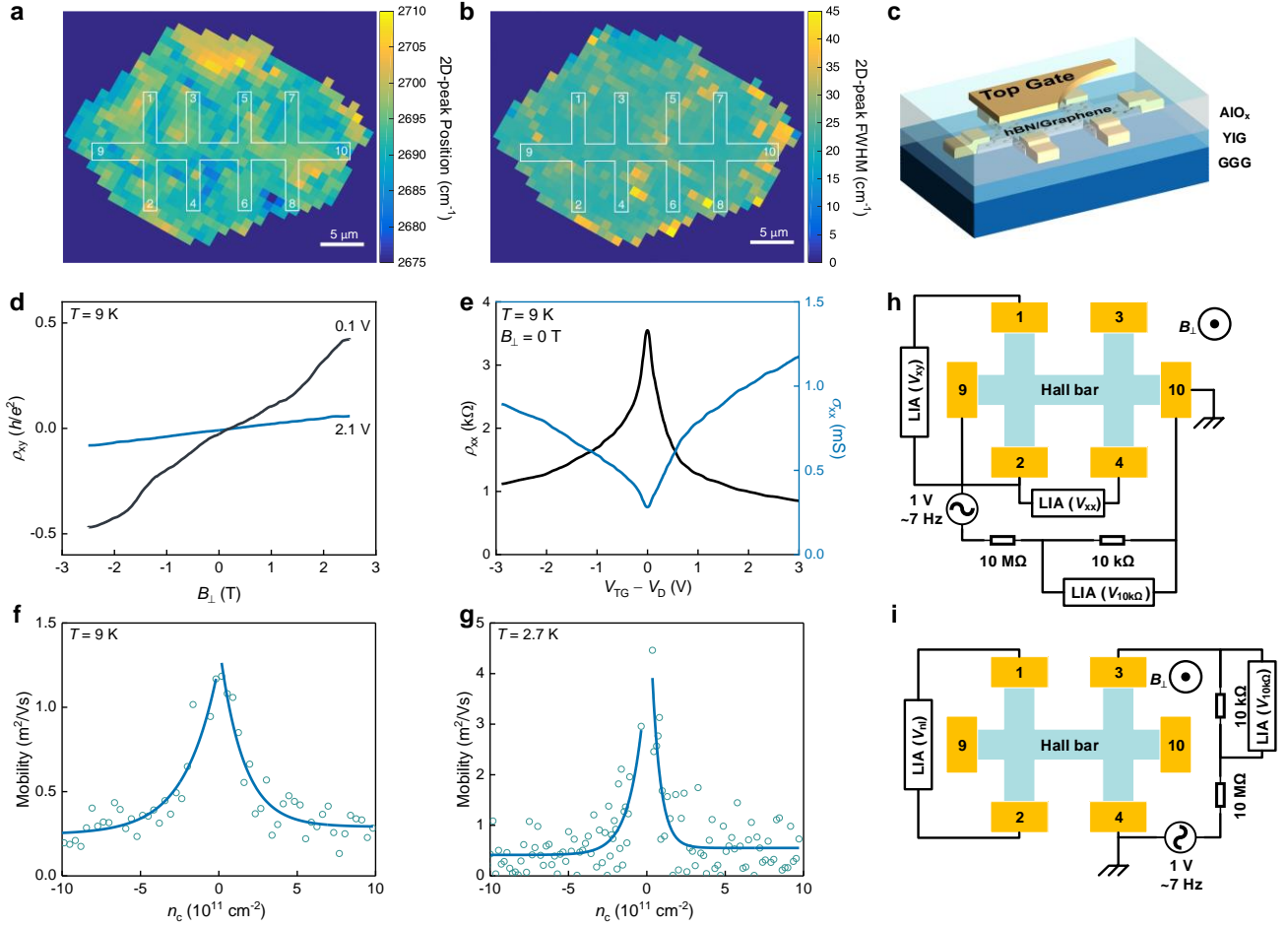

FIG. S4. Raman spectra and electrical properties of hBN/graphene Hall bars on YIG. (a)-(b) Raman spectra maps of (a) positions (cm$^{-1}$) and (b) FWHM (cm$^{-1}$) of the 2D-peak at 293 K, where white solid lines show the position of the Hall bar. (c) Schematic illustration of a Hall bar device. (d) Magnetic field ($B_\perp$) dependence of Hall resistivity ($\rho_{xy}$) at different gate voltage at 9 K. (e) Gate-voltage-dependence of longitudinal resistivity ($\rho_{xx}$) and conductivity ($\sigma_{xx}$) in 0 T at 9 K. (f) and (g) Field-effect mobility vs gate voltage at 9 K and 2.7 K. (h) $R_{xx}$ and $R_{xy}$ measurement setup, (i) $R_{nl}$ measurement setup. LIA is the lock-in amplifier.

## S5. Ohmic contribution to $R_{nl}$

Several sources may induce nonlocal resistance in the absence of a magnetic field. One source is the ohmic effect [1], which is given by

$$R_{\text{nl},\Omega} = \frac{W}{\pi L} R_{xx} \ln\left[\frac{\cosh(\pi L/W)+1}{\cosh(\pi L/W)-1}\right], \tag{S1}$$

where $L$ and $W$ are the channel length and width as graphically defined in the left inset of Fig. 2(b). In zero magnetic field for the hBN/graphene Hall bar on YIG, $L/W$ = 2.75 and $R_{xx}$ = 9.8 kΩ, and from Eq. (S1) we find $R_{\text{nl},\Omega} \approx 1$ Ω, which is two orders of magnitude smaller than the measured $R_{\text{nl}}$ [Fig. S5(a)]. Furthermore, $R_{\text{nl}}$ is sharper than $R_{xx}$ when top-gate voltage approaches the Dirac point ($V_D$). We can conclude that $R_{\text{nl}}$ is not proportional to $R_{xx}$ in zero magnetic field meaning that the nonlocal resistance is dominated from other factors than the ohmic effect.

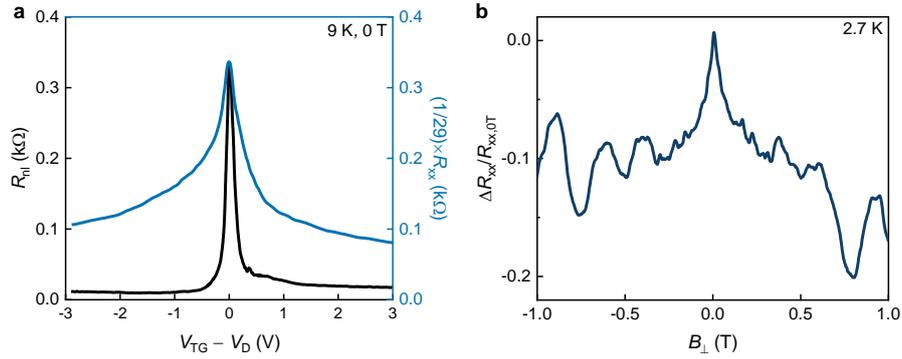

FIG. S5. Ohmic contribution to $R_{\text{nl}}$ and weak localization effect for hBN/graphene Hall bar on YIG. (a) Scaled $R_{xx}$ and $R_{\text{nl}}$ in zero magnetic field at 9 K. (b) Low magnetic field normalized magnetoresistance $(R_{xx} - R_{xx,0T})/R_{xx,0T}$ vs $B_\perp$ at 2.7 K. It shows a sharp and symmetric peak around zero magnetic field, arising due to the weak localization effect.

## S6. Magnetoresistance in low magnetic fields

In low magnetic fields our device exhibits a sharp negative magnetoresistance which is attributed to the weak localization effect [Fig. S5(b)]. The device shows a high quality which is confirmed by Shubnikov-de Haas oscillations appearing in magnetic field as low as 1.0 T (Fig. S6). As Rashba spin-orbit coupling in graphene should give rise to the weak antilocalization effect, there is no strong spin-orbit coupling in our device.

## S7. Thermal contribution to $R_{\text{nl}}$

In all transport measurements, we use an alternating-current excitation of 50 nA at 2.7 K and 100 nA at 9 K with a frequency of 7 Hz. These low current amplitudes are chosen to minimize thermal contribution to the nonlocal transport due to Joule heating and Ettingshausen effects whilst simultaneously maximizing the signal-to-noise ratio of the measured voltages.

Joule heating can give rise to the second harmonic nonlocal signal $R^{2f}_{\text{nl},J}$, and Ettingshausen effect can lead to the first harmonic nonlocal signal $R^{f}_{\text{nl},E}$ [2,3]. For all the Hall bars, we measure both the first and second harmonic nonlocal signals using lock-in amplifiers. In 2.5 T, $R_{nl,J}^{2f}$ is typically two orders of magnitude below the $R_{nl}$ [Figs. S7(a)-S7(b)] and in the 8-10 Ω range when $I$ = 100 nA. Figures S7(c)-S7(d) show that $R_{nl,J}^{2f}$ is directly proportional to the excitation current as expected and hence the excitation current is always below 100 nA during the measurements to minimize the thermal contribution. In addition, when the current and voltage probes are switched to reverse the excitation current direction, $R_{nl,J}^{2f}$ changes sign due to reverse heat flow along the Hall bar [Figs. S7(c)-S7(d)]. In zero magnetic field, $R_{nl,J}^{2f}$ is less than 1 Ω, which is negligible shown in Fig. S7(e).

Ettingshausen contribution is due to the heat flow generated by Nernst-Ettingshausen effect, which can be described as $R_{nl,E}^{f} \propto S_{yx}^{2} T$ ($S_{yx}$, transverse thermopower coefficient and $T$, temperature) [3]. The maximum $R_{nl,E}^{f}$ and $S_{yx}$ both occur at $N$ = 0 Landau level, but $S_{yx}$ changes sign when the gate voltage locates between adjacent Landau levels or in the center of a Landau level [2], in which $R_{nl,E}^{f}$ should have peaks. Figure S7(f) compares the $R_{nl}$ and $R_{xx}$ as a function of top-gate voltage, but $R_{nl}$ does not show other peaks except for at the Landau level positions, which indicates the Ettingshausen effect contribution is negligible.

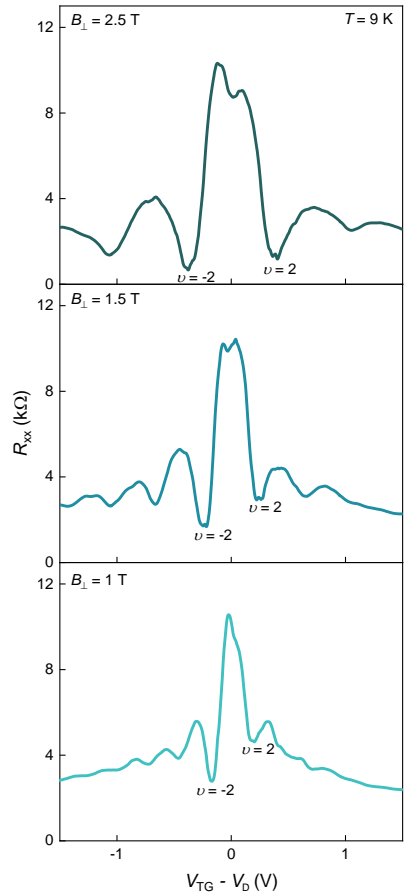

FIG. S6. $R_{xx}$ vs $V_{TG}$ - $V_D$ for $B_\perp$ of 1 T, 1.5 T and 2.5 T for device I at 9 K (labelled).

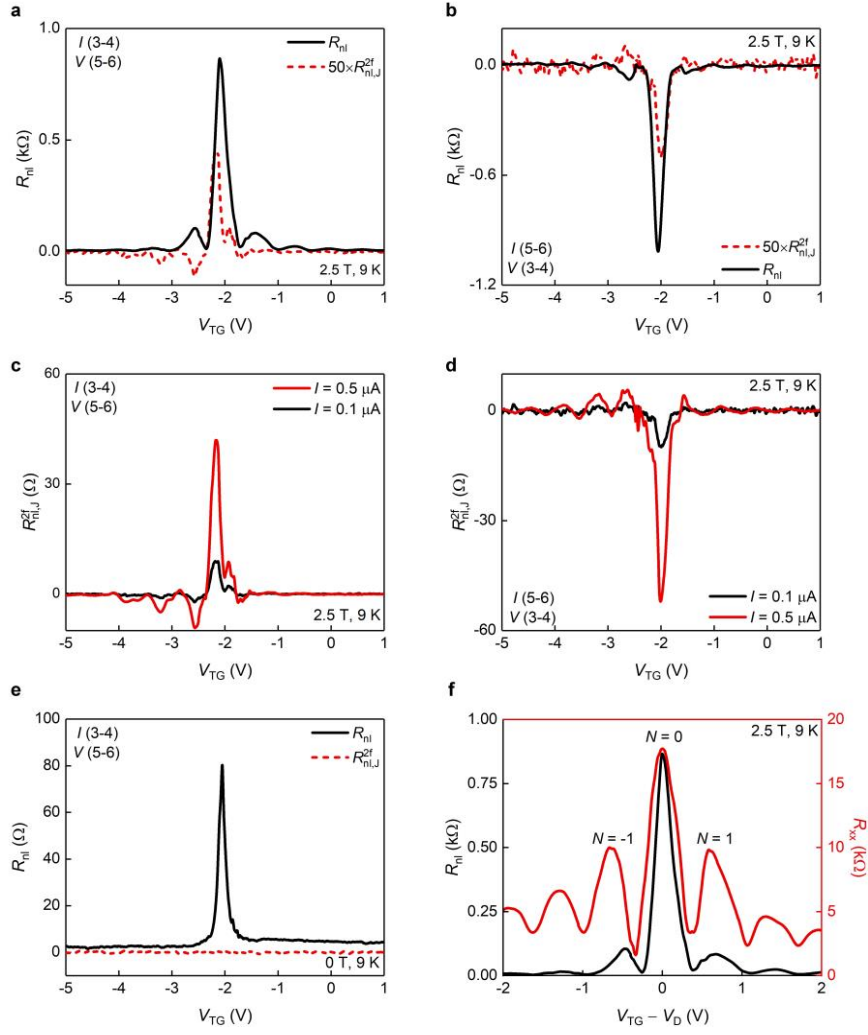

FIG. S7. Comparison of the first and the second harmonic signals of $R_{nl}$ and $R_{xx}$ in 2.5 T and 0 T for device I. (a) $R_{nl,J}^{2f}$ (multiplied by 50) and (b) $R_{nl}$ vs $V_{TG}$ in 2.5 T with reverse nonlocal connections. (c)-(d) $R_{nl,J}^{2f}$ vs $V_{TG}$ when inject current $I$ = 0.1 μA and 0.5 μA with different connections. (e) Comparison of the $R_{nl,J}^{2f}$ and $R_{nl}$ in 0 T. (f) Comparison of $R_{nl}$ and $R_{xx}$ shown no additional oscillations in $R_{nl}$ at $N$ = ±1 Landau levels, indicating that the Ettingshausen effect is negligible. All data are recorded at 9 K.

## S8. Nonlocal and local transport measurements on hBN/graphene/AlO$_x$/YIG control Hall bar

We investigate $R_{nl}$ and $R_{xx}$ of hBN/graphene/AlO$_x$/YIG control Hall bar in which the graphene is decoupled from YIG by a thin layer of AlO$_x$ (6 nm). The Hall-effect mobility is 15,752 cm$^2$ V$^{-1}$ s$^{-1}$ with 5×10$^{11}$ cm$^{-2}$ carrier density at 9 K, and the field-effect mobility is in the range of 12,000 – 20,000 cm$^2$ V$^{-1}$ s$^{-1}$ with a carrier density between 10$^{11}$ and 10$^{12}$ cm$^{-2}$. For $B_\perp$ = 12 T, both $\sigma_{xx}$ and $R_{nl}$ show a single peak at the Dirac point [Fig. S8(a)], and the Hall conductivity ($\sigma_{xy}$) vs top-gate voltage only shows plateaus corresponding to $v$ = ±2 [Fig. S8(b)]. In 12 T and 0 T, $R_{nl,J}^{2f}$ is around two orders of magnitude below $R_{nl}$ [Figs. S8(c)-S8(d)]. In zero magnetic field for the

hBN/graphene/AlO$_x$/YIG, $L/W$ = 2.3 and $R_{xx}$ = 10 kΩ, and from Eq. (S1) we find $R_{nl,\Omega} \approx 4$ Ω, which is one order of magnitude smaller than the measured $R_{nl}$ [Fig. S8(e)].

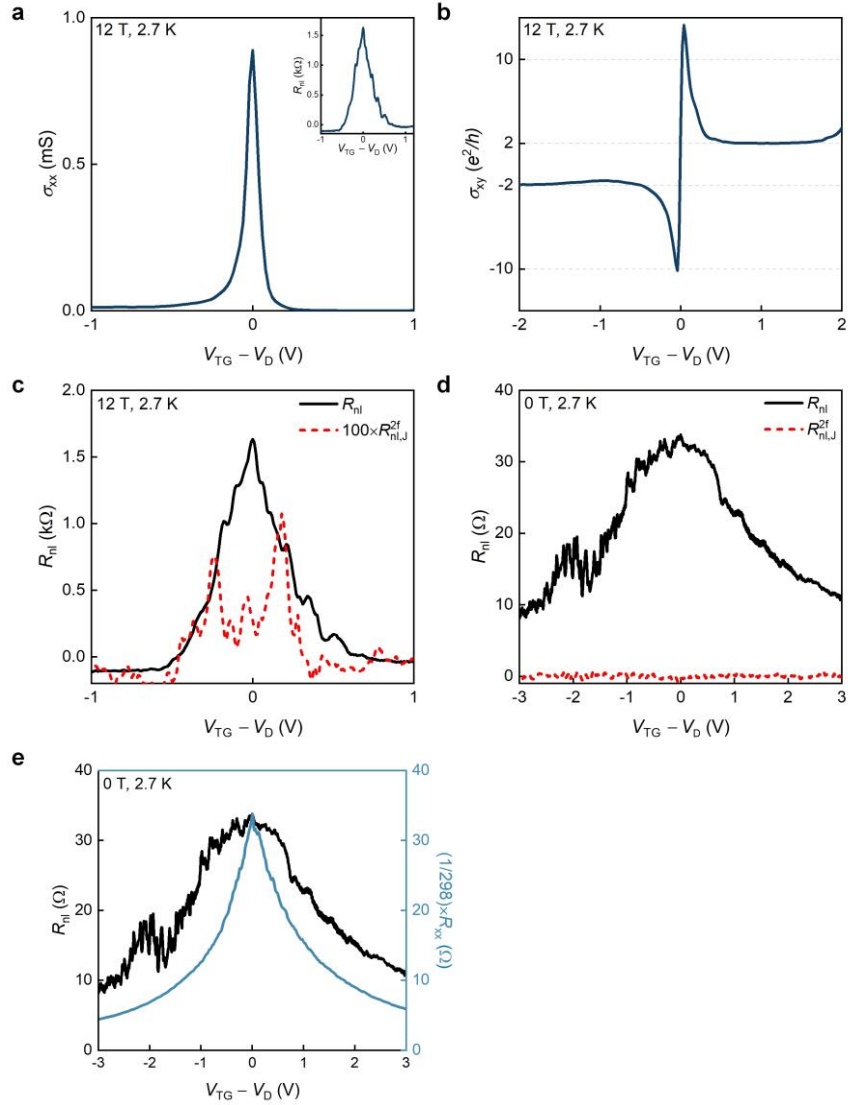

FIG. S8. Transport measurements for hBN/graphene/AlO$_x$/YIG Hall bar. (a) $\sigma_{xx}$ vs $V_{TG} - V_D$ with $B_\perp$ = 12 T. The inset shows $R_{nl}$ vs $V_{TG} - V_D$. (b) $\sigma_{xy}$ vs $V_{TG} - V_D$ with $B_\perp$ = 12 T, which only shows the plateaus of conductance corresponding to $\nu = \pm 2$. (c) Comparison between $R_{nl,J}^{2f}$ (multiplied by 100) and $R_{nl}$ with $B_\perp$ = 12 T. (d) Comparison between $R_{nl,J}^{2f}$ and $R_{nl}$ in zero magnetic field. (e) Comparison between the scaled $R_{xx}$ and $R_{nl}$ in zero magnetic field. All data are recorded at 2.7 K.

## S9. Anomalous Hall effect in hBN/graphene/YIG Hall bar

In the low applied perpendicular magnetic field range ($B_\perp$ < 1 T), Hall resistivity ($\rho_{xy}$) of hBN/graphene/YIG Hall bar shows nonlinear relation with $B_\perp$ in Fig. S9(a). According to an empirical relation [14,15] between $\rho_{xy}$, $B_\perp$, and magnetization ($M$), $\rho_{xy} = R_H(B_\perp) + R_{AHE}(M)$. $R_H(B_\perp)$ represents the linear ordinary Hall effect due to Lorentz force. $R_{AHE}(M)$ represents the anomalous Hall effect (AHE) due to the spontaneous magnetization. We remove the linear ordinary Hall effect background from measured $\rho_{xy}$ and get anomalous Hall resistance ($R_{AHE}$). In Figs.

S9(b) and S9(c), $R_{AHE}$ saturates at around 0.2 T which is consistent with the maximum out-of-plane magnetic field required to fully magnetize YIG in Fig. S3(d). The largest $R_{AHE}$ of around 250 Ω is observed at 9 K with a carrier density of $2.0\times10^{11}$ cm$^{-2}$. When gate voltage is changing away from Dirac point, $R_{AHE}$ decreases in Fig. S9(b). Figure S9(c) shows temperature dependence of $R_{AHE}$, and $R_{AHE}$ decreases with increasing temperature. The AHE could arise from induced magnetic exchange field and a reasonably weak spin-orbit coupling effect. In hBN/graphene/AlO$_x$/YIG control Hall bars, there is only linear ordinary Hall effect.

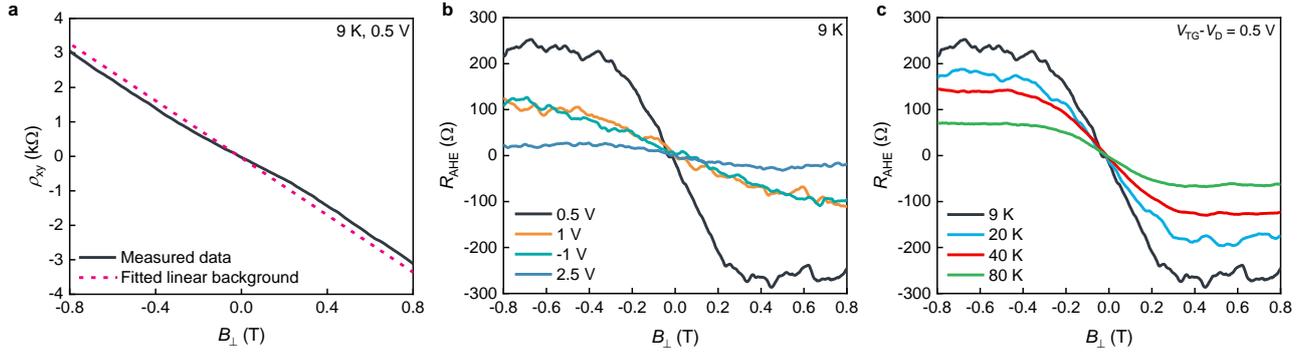

FIG. S9. Anomalous Hall effect in device I. (a) $\rho_{xy}$ vs $B_\perp$ at 9 K and $V_{TG} - V_D$ = 0.5 V, dotted line shows the fitted linear Hall effect background. (b) Anomalous Hall resistance ($R_{AHE}$) vs $B_\perp$ with different gate voltages ($V_{TG} - V_D$) at 9 K. (c) $R_{AHE}$ vs $B_\perp$ with different temperatures at $V_{TG} - V_D$ = 0.5 V.

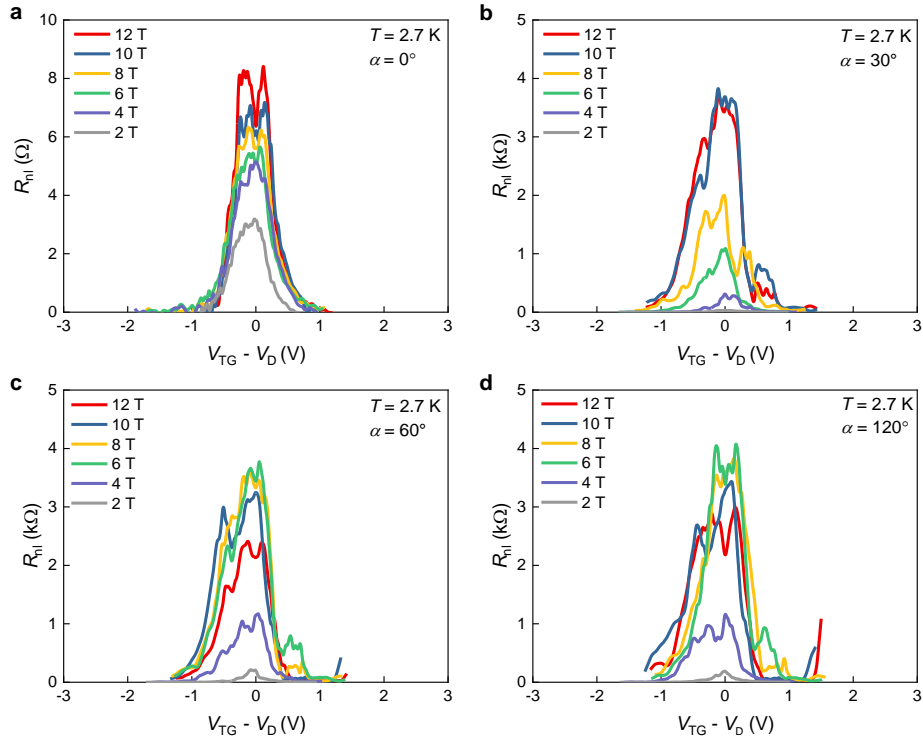

FIG. S10. $R_{nl}$ vs $V_{TG} - V_D$ for different applied magnetic fields (labelled) with rotation angle ($\alpha$) of 0°, 30°, 60° and 120° for device II at 2.7 K.

**S10. Nonlocal transport measurement with rotation angle on hBN/graphene/YIG Hall bar (device II)**

In order to confirm the magnetic states at $v$ = 0, nonlocal measurements with applied rotating magnetic field from $B_\perp$ ($\alpha$ = 90º) to $B_{//}$ ($\alpha$ = 0º) are performed on hBN/graphene Hall bar on YIG (device II). The measurement results are shown in Fig. 5 and Fig. S10. When $\alpha$ = 0º, $R_{nl}(\alpha = 0º)$ for $B$ = 2−12 T are less than 10 Ω, which is three orders of magnitude smaller than $R_{nl}(\alpha = 90º)$. Because when $B$ is in-plane, the hBN/graphene Hall bar on YIG is not in the quantum Hall regime and does not have Landau levels, then the bulk state is not gapped and there is no edge modes which would give rise to the nonlocal resistance. Under these conditions the mechanism for the nonlocal resistance at $\alpha$ = 0º is very different. It may arise due to the Zeeman spin Hall effect, but this effect is expected to be very weak.

**S11. Theoretical description of the competition between F and CAF states**

We discuss the transitions between F and CAF states in the light of the theory [4,5] and the previous experiment [6]. The dependence of the sublattice spin tilting angle in the CAF state on the applied magnetic field (**B**) and the YIG induced magnetic exchange field (**M**$_{ex}$) is estimated.

In the non-interacting limit (when Zeeman field and spin-orbit interactions are neglected), graphene supports four zero-energy Landau levels and has spin degeneracy as well as valley (K, K') degeneracy. In each valley the wavefunctions reside on one of the sublattices (A, B), the valley index is directly related to the degree of freedom of sublattice (K ↔ A, K' ↔ B). The electron-electron and electron-phonon interactions break the valley symmetry on the lattice scale and the generated valley anisotropy determines the magnetic state of the $v$ = 0 in graphene. In a systematic theoretical study of the possible anisotropy, there are a large number of different possible symmetry-broken states [4], but from the experiment results in Refs. [6–8] it indicates that the interactions lead to an AF state (both bulk and edge modes are gapped) with opposite spin-polarization on the two sublattices. Moreover, the applied magnetic field and the magnetic proximity effect of YIG lead to the breaking SU(4) symmetry. The Hamiltonian of total magnetic field (**M**$_T$, the sum of the Zeeman field and **M**$_{ex}$ induced by the YIG) is described as

$$H_M = -\mu_B \boldsymbol{M}_T \cdot \boldsymbol{\sigma}, \boldsymbol{M}_T = \frac{g}{2}\boldsymbol{B} + \boldsymbol{M}_{ex}. \tag{S2}$$

The sublattice spins tend to be parallel to **M**$_T$. All **M**$_T$, **B** and **M**$_{ex}$ are in unit of tesla. $\mu_B$ is Bohr magneton, $g$ is gyromagnetic ratio and $\boldsymbol{\sigma}$ are the Pauli matrices.

We have made the following assumptions: (1) $g$ = 2; (2) **M**$_T$ is spatially uniform, and **M**$_{ex}$ is considered as a spatial average of the magnetic exchange field induced by the YIG in graphene; (3) There is a disordered interface between graphene and YIG, so that the sublattice (valley) symmetry is not broken on average; (4)

$$\boldsymbol{M}_{\mathbf{ex}} = M_{\mathrm{ex}} \frac{\boldsymbol{B}}{\|\boldsymbol{B}\|}, \tag{S3}$$

in the absence of spin-orbit coupling effect, $\boldsymbol{M_{ex}}$ is parallel or antiparallel to $\boldsymbol{B}$, but due to the ferrimagnetic nature of YIG, the magnitude of $\boldsymbol{M_{ex}}$ may depend on the direction of $\boldsymbol{B}$. However, if the locations of the oppositely polarized magnetic moments do not depend on the direction of $\boldsymbol{B}$, this dependence is expected to be weak. Therefore the magnitude of the $\boldsymbol{M_{ex}}$ is independent of the direction of $\boldsymbol{B}$; (5) The magnitude of the $\boldsymbol{M_{ex}}$ does not depend on $||\boldsymbol{B}||$, when $\boldsymbol{B}$ is sufficiently large so that the magnetization of YIG is saturated. In addition, in-plane and out-of-plane anisotropies of YIG are not important.

The competition between the AF and F states leads to the CAF state where spins in the two sublattices not only tilt along the direction of the $\boldsymbol{M_T}$, but also they have components perpendicular to $\boldsymbol{M_T}$ which point in opposite directions (preferred by the electron-electron interactions leading to an AF state). The spin-directions in the two sublattices can be calculated by minimizing the energy [4,5]

$$E(\theta) = E_0 + \frac{A_s}{\pi l_B^2}[-\mu_B M_T(B)\cos\theta + u(B_\perp)\cos^2\theta], \tag{S4}$$

where $E_0$ is a constant, $A_s$ is the area of the sample and $l_B$ is the magnetic length. The anisotropy energy $u$ depends on the microscopic interactions that break the valley symmetry on the lattice scale, and on the Landau level wavefunctions, therefore $u$ can be controlled with out-of-plane mangetic field ($B_\perp$).

Under the assumptions discussed above, the magnitude of total applied magnetic field [$||\boldsymbol{M_T}||$ = $M_T(B)$] depends on the magnitude of the applied magnetic field ($B = ||\boldsymbol{B}||$). The spin-directions relative to the direction of $\boldsymbol{M_T}$ are described by the polar angle $\vartheta$, and the azimuthal angles ($\varphi$, $\varphi + \pi$) in the two sublattices. The energy does not depend on the azimuthal angle, so $\varphi$ describes spontaneously broken U(1)-symmetry in the CAF state. The AF state is achieved when $\boldsymbol{M_T}$ = 0 and in this case $\vartheta = \pi/2$, indicating that the spins in the two sublattices point in opposite directions. When $\boldsymbol{M_T}$ = 0, $\vartheta$ can be relative to any directions and the AF state is described by a spontaneously broken SU(2)-symmetry. In the F state, the magnitude of $\boldsymbol{M_T}$ is so large that spins are fully polarized along the direction of $\boldsymbol{M_T}$, then there is no spontaneously broken symmetry in this case (paramagnetic phase). The angle $\vartheta$ is calculated by

$$\cos\theta = \begin{cases} 1, & \mu_B M_T > 2u(B_\perp) \\ \mu_B M_T / 2u(B_\perp), & \mu_B M_T < 2u(B_\perp) \end{cases}, \tag{S5}$$

so that the transition from F to CAF state occurs at

$$\mu_B M_c = 2u(B_\perp). \tag{S6}$$

The anisotropy energy has not been described as an equation theoretically, therefore we determine it by utilizing earlier experiment results [6]. In [6] the edge mode conductance is measured as a function of $B$ for various values of $B_\perp$. In the CAF state, the edge mode is gapped at the Dirac point giving a zero conductance, while the F state supports counter-propagating edge modes. If they are ballistic (i.e. the length of the system is shorter than the mean free path), the conductance in a simple two-terminal geometry is given by $G = 2e^2/h$. Indeed, experimentally the conductance $G(B, B_\perp)$ shows a sharp crossover from 0 to $2e^2/h$. Then the critical applied magnetic field $B_c$ can be estimated by requiring

$$G(B_c, B_\perp) = e^2/h. \tag{S7}$$

As in [6] the magnetic exchange field is zero, we get $M_c = B_c$ and $u$ can be determined from Eq. (S6). The phase transition lines in the ($M_T$, $B_\perp$)-plane for two samples reported in [6] are shown in Fig. S11(a). By averaging the two lines, we have

$$u(B_\perp) = 5\mu_B(B_\perp + 0.5), \tag{S8}$$

which is a reasonable estimation in the light of Refs. [4,5]. But the shift of 0.5 T indicates that a nonlinear function is approximately linearized. In the experiment of Ref. [9], the thermal activation gap increases approximately linearly with $B_\perp$ up to 30 T, so the assumption of linear dependence is reasonable.

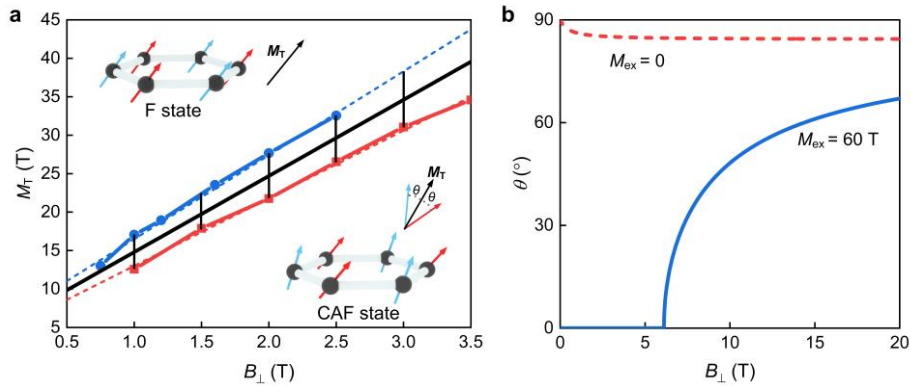

FIG. S11. Phase transition diagrams. (a) $M_T$ vs $B_\perp$ for graphene in which the solid and dashed blue lines ($M_T \approx 10.9 B_\perp + 5.6$), and solid and dashed red lines ($M_T \approx 8.8 B_\perp + 4.2$) are calculated from [6] using extracted data in blue and red dots. The solid black line ($M_T \approx 9.9 B_\perp + 4.9$) is the average of the blue and red lines. (b) The polar angle $\vartheta$ as a function of $B_\perp$. The solid blue (dashed red) line corresponds to $M_{ex} = 60$ T ($M_{ex} = 0$). In the presence of magnetic exchange field, it is able to control $\vartheta$ with $B_\perp$. But in the absence of magnetic exchange field, $\vartheta \approx \pi/2$ for all values of $B_\perp$ approximately corresponding to an AF state.

According to the estimation for $u(B_\perp)$, we find that $2\vartheta = 0.93\pi$ in the absence of magnetic exchange field or for extremely large $B_\perp$. It means that the spins practically point in opposite directions, leading to an AF state, which is consistent with the results that $\nu = 0$ state in graphene is approximately spin-unpolarized [9] as well as both bulk and edge modes are gapped [9,10]. In the case of tilted magnetic field (**B**), it leads to a CAF state and $\vartheta$ decreases with increasing **B** when $B_\perp$ is constant. To achieve a transition from F to CAF state, it needs a very large magnetic field (> 15 T) [6].

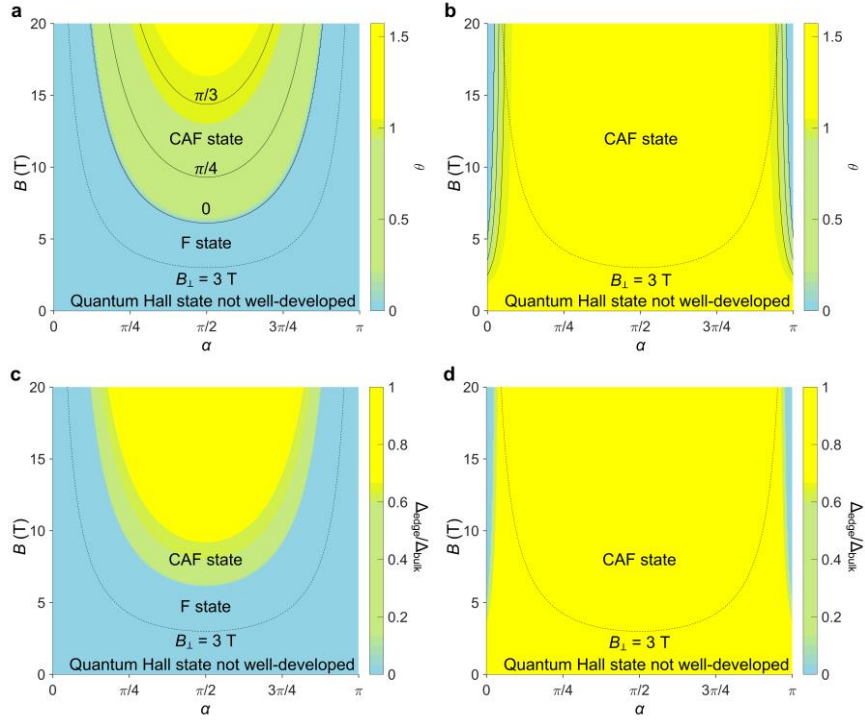

FIG. S12. Phase transitions vs the magnitude and direction ($\alpha$) of magnetic field. (a)-(b) The polar angle $\vartheta$ as a function of $B$ and the angle $\alpha$ for (a) $M_{ex}$ = 60 T and (b) $M_{ex}$ = 0. (c)-(d) The energy gap of the edge modes $\Delta_{edge}$ as a function of $B$ and the angle $\alpha$ for (c) $M_{ex}$ = 60 T and (d) $M_{ex}$ = 0. The dashed lines in all figures indicate $B_\perp$ = 3 T. Below this line the quantum Hall state is not well-developed in our samples.

Our experiments indicate that the magnetic proximity effect from YIG does not lead to strong orbital effects in graphene, and therefore the estimation for the anisotropy energy in Eq. (S8) is valid. In Fig. 5, the transitions from F to CAF state occur in $B_\perp \approx$ 6 T. So the magnitude of $\boldsymbol{M_{ex}}$ in graphene induced by YIG is estimated to be 60 T. Moreover we calculate how the angle $\vartheta$ depends on the magnitude of $B_\perp$. Interestingly, $\vartheta$ can be modulated over a wide range starting with a relatively low applied magnetic field (> 6 T) in Fig. S11(b). Such kind of control of the magnetic state is not possible in the absence of magnetic exchange field.

The polar angle $\vartheta$ can be varied by rotating the direction of the magnetic field. Figures S12(a)-S12(b) plot the $\vartheta$ as a function of $B$ and the $\alpha$ ($B_\perp = B\sin\alpha$ and $B_\parallel = B\cos\alpha$). In the presence of magnetic exchange field induced by YIG [Fig. S12(a)], it is possible to control $\vartheta$ with the direction and magnitude of $\boldsymbol{B}$ leading to a transition between CAF state and F state. If $B$ is constant, varying the angle $\alpha$ from $\pi/2$ to 0 firstly causes a transition from CAF to F state, and then by further decreasing $\alpha$ the system is driven away from the quantum Hall regime because $B_\perp$ decreases. We expect that the quantum Hall state is not well-developed in our samples for magnetic fields below 3 T. In the absence of magnetic exchange field [Fig. S12(b)], it requires a large $B$ in order to access the regime where $\vartheta$ can be changed substantially. And when magnetic fields are in the order of 6-10 T, the angle $\vartheta$ remains close to $\pi/2$ (approximately an AF state) until $\alpha$ is so small that the system is driven out of the quantum Hall regime.

It is worth pointing out that magnetic exchange field lowers the energy of both the CAF and F states. In the CAF state, it can be described as $-\delta E \propto M_T^2/u(B_\perp)$ and as $-\delta E \propto M_T$ in the F state. Therefore, magnetic exchange field can lead to both CAF and F states in samples where the $v = 0$ quantum Hall state would not be realized in the absence of magnetic exchange field. This explains why the $v = 0$ quantum Hall state is only observed in the sample where the coupling to YIG is present and not in control samples where magnetic exchange field is absent.

## S12. Energy gap of the edge modes

The low-energy edge excitations in this type of quantum Hall systems are collective [11–13]. While the properties of these collective excitations are different from the single-particle excitations, an estimation of the energy gap of the edge modes can be obtained using a simplified mean-field single-particle approach proposed in Ref. [5].

In the AF state, the bulk and edge modes are gapped, whereas in the F state the bulk mode are gapped but gapless counter-propagating modes (protected by spin-rotation symmetry) appear at the edge. As the CAF state is a mixture between the AF state ($\vartheta = \pi/2$) and F state ($\vartheta = 0$), the edge gap has to gradually decrease when $\vartheta$ decreases from $\pi/2$ to 0. This behavior is captured by the mean-field Hamiltonian [5]

$$H = -\xi(k_x)\tau_x\sigma_0 - (\mu_B M_T + \Delta_z)\tau_0\sigma_z - \Delta_x\tau_z\sigma_x, \quad \text{(S9)}$$

where the Pauli matrices $\tau_i$ and $\sigma_i$ operate on the valley and spin degrees of freedom, respectively, and the spin quantization axis (z-direction in the spin space) has been chosen to be along the direction of the applied magnetic field. $\Delta_x$ and $\Delta_z$ are the Hartree-Fock mean-field potentials due to electron-electron interacitons. Here $M_T = ||\boldsymbol{M_T}||$, $\xi(k_x)$ describes the dispersion of the Landau levels and $k_x$ is the momentum along the direction of the edge (x-direction), which is related to the position $y_{k_x} = k_x l_B^2$ perpendicular to the edge in Landau gauge. When $y_{k_x}$ is deep inside the bulk, $\xi(k_x) = 0$, and when $y_{k_x}$ approaches the edge, $\xi(k_x)$ increases. In addition, we have introduced the mean-field potentials obtained by decoupling of the interactions,

$$\Delta_z = 2V_z(B_\perp)\cos\theta, \Delta_x = 2V_x(B_\perp)\sin\theta. \quad \text{(S10)}$$

The mean-field potential $\Delta_z$ arises due to the spin-component parallel to $\boldsymbol{M_T}$, and $\Delta_x$ is due to the spin-component perpendicular to $\boldsymbol{M_T}$. We have neglected the spatial dependence of the mean-field potentials $\Delta_z$ and $\Delta_x$. The effective interaction strengths $V_z(B_\perp)$ and $V_x(B_\perp)$ increase with increasing $B_\perp$ and the related anisotropy energy is

$$u(B_\perp) = V_x(B_\perp) - V_z(B_\perp). \quad \text{(S11)}$$

Moreover, $V_x(B_\perp)$ is directly related to the bulk gap $\Delta_{\text{bulk}}$ in the AF and CAF states,

$$\Delta_{\text{bulk}} = 4V_x(B_\perp). \tag{S12}$$

Therefore $V_x(B_\perp)$ could be estimated using the thermal activation gap, and $V_z(B_\perp)$ could be determined using Eqs. (S8) and (S11). In the experiment of Ref. [9], the thermal activation gap increases approximately linearly with $B_\perp$, so that all interaction strengths $V_x(B_\perp)$. $V_z(B_\perp)$ and $u(B_\perp)$ depend linearly on $B_\perp$ in the range of magnetic fields we consider here. However, in the following we describe the energy gap of the edge mode $\Delta_{\text{edge}}$ in the unit of $\Delta_{\text{bulk}}$, so the exact values of the interaction strengths $V_x(B_\perp)$ and $V_z(B_\perp)$ are not required.

According to the energy spectrum from the Hamiltonian in Eq. (S9), the energy gap of the edge mode is given by

$$\Delta_{\text{edge}} = \begin{cases} 0, & \mu_B M_T > 2u(B_\perp) \quad \text{(F state)} \\ \Delta_{\text{bulk}} \sin\theta, & \mu_B M_T < 2u(B_\perp) \, \text{(CAF state)} \end{cases}. \tag{S13}$$

$\Delta_{\text{edge}} = \Delta_{\text{bulk}}$ in the AF state ($\vartheta = \pi/2$), and gap vanishes ($\Delta_{\text{edge}} \to 0$) when it approaches the F state ($\vartheta \to 0$). Furthermore, in the presence of magnetic exchange field [Fig. S12(c)], it is able to tune $\Delta_{\text{edge}}$ efficiently using relatively low applied magnetic fields. But in the absence of magnetic exchange field [Fig. S12(d)], the energy gap satisfies $\Delta_{\text{edge}} \approx \Delta_{\text{bulk}}$ unless very large magnetic fields are applied.